\documentclass[11pt]{article}

\usepackage[margin=1.1in]{geometry}
\usepackage[utf8]{inputenc}
\usepackage[T1]{fontenc}
\usepackage{lmodern}
\usepackage{amsmath,amssymb}
\usepackage{graphicx}
\usepackage{booktabs}
\usepackage{tabularx}
\usepackage[round]{natbib}
\usepackage{url}
\usepackage[hidelinks]{hyperref}
\usepackage[activate={true,nocompatibility},final,tracking=false,kerning=false,spacing=false,factor=1100]{microtype}
\microtypesetup{expansion=false}

\newcolumntype{Y}{>{\raggedright\arraybackslash}X}

\title{\bfseries Epistemic Transfer in AI-Assisted Verification:\\A Framework and Evaluation Protocol}

\author{Christoph Trattner\\[2pt]
\normalsize SFI MediaFutures, Research Centre for Responsible Media Technology and Innovation\\
\normalsize Department of Information Science and Media Studies, University of Bergen, Norway\\
\normalsize \texttt{christoph.trattner@uib.no} \,\textbullet\, ORCID: 0000-0002-1193-0508}

\date{September 2026\\[4pt]\normalsize Revised working paper. Comments welcome.}

\begin{document}

\maketitle

\begin{abstract}
\noindent AI tools can improve claim judgments while leaving open what users can do later without them. This paper develops an evaluation framework for \emph{epistemic transfer}: the effect of prior AI-assisted verification on delayed judgments of novel claims under a specified access regime. The contribution is a verification-specific synthesis of learning, transfer, and human--AI evaluation, organized around two complementary estimands. The Epistemic Transfer Effect (ETE) compares delayed performance after alternative practice conditions. Tool-Removal Cost (TRC) compares immediate performance with and without assistance after practice; despite its name, it measures a current availability effect, not skill loss or psychological dependence. The proposed randomized protocol includes answer-first and evidence-first interfaces, active practice, a no-additional-practice comparator, and held-out claims. It specifies how to account for learning opportunities introduced by assessment, elicit confidence probabilities, average model predictions over a target population, and handle attrition and uncertainty. Reading ETE and TRC together distinguishes relative capability gains, equivalence, transfer penalties, and unresolved outcomes. A ``verification-on-loan'' profile is explicitly comparator-relative and cannot be inferred from a nonsignificant delayed contrast. A brief illustration from a two-wave verification study shows why these distinctions matter: an uncertain delayed interface contrast and an ordered assisted--unassisted probe cannot establish a clean transfer profile. The framework makes a practical demand: when independent judgment matters, evaluate both what assistance contributes now and what prior use changes later.

\end{abstract}

\section{Introduction}

AI systems increasingly help people judge whether online information is true, misleading, or unreliable. Some systems retrieve evidence and predict claim veracity \citep{guo2022survey}. Others summarize sources and produce fluent, natural-language verdicts. Platforms also add labels, provenance signals, or other credibility cues. Most evaluations of these systems focus on what happens at the moment of use: whether the tool is accurate, whether users follow good advice, or whether the human--AI pair performs better than the human or the AI alone \citep{bansal2021does,vaccaro2024combinations}.

Those are important outcomes, but they are not the whole story. In many settings, the bigger question is what repeated use of such tools does to the user. A helpful system may expose people to better evidence and better strategies. But it may also reduce the need to search, compare sources, inspect uncertainty, or form an independent judgment. In other words, the same system can improve performance now while having positive, null, or negative effects on later independent performance.

A second distinction is also important: an interface feature is not the same thing as the action a user takes with it, and neither is identical to the user's psychological experience of agency or control. The HAII--TIME framework distinguishes a \emph{cue route}, in which visible AI or interface attributes trigger heuristics, from an \emph{action route}, in which users engage the action possibilities provided by the system \citep{sundar2020machine}. Recent work on conversational generative AI makes this distinction concrete: a verification affordance can be presented merely as a cue, required as an action, or left for voluntary use, with different consequences for trust and verification behavior \citep{liao2026chat}. For transfer research, this means that later capability should be treated as a downstream outcome of an interaction architecture, while enacted verification, perceived agency, effort, anchoring, and trust remain candidate mechanisms rather than interchangeable labels.

This issue is familiar in nearby literatures. Learning research distinguishes doing well during supported practice from actually learning something that remains after support is removed \citep{soderstrom2015learning,roediger2006test}. Cognitive-offloading research shows that people shift effort and memory when external resources are available \citep{risko2016cognitive,fisher2015searching,storm2015saving}. Automation research shows that high levels of support can reduce opportunities to maintain manual or cognitive skills \citep{bainbridge1983ironies,casner2014retention,onnasch2014human,parasuraman1997humans}. Recent AI studies suggest that the same concern matters for generative and clinical AI systems as well \citep{bastani2025generative,budzyn2025endoscopist,deverna2024fact,rani2026dialogues,liu2026ai,shen2026how,stromberg2026penalty}.

These literatures already establish the distinction between supported performance and learning. The remaining methodological task is to translate that distinction into a shared protocol with explicit comparators for AI-assisted verification. This paper offers that synthesis; it does not claim that delayed transfer or tool withdrawal is a new experimental idea.

\subsection{Objective and Contributions}

The goal of this paper is to offer a practical framework for studying what people retain from AI-assisted verification. I make three contributions:

\begin{enumerate}
\item I specify epistemic transfer for claim verification and distinguish it from immediate correction, trust, reliance, and human--AI team performance.
\item I define two comparator- and regime-specific quantities: the Epistemic Transfer Effect (ETE) for delayed performance differences and Tool-Removal Cost (TRC) for the immediate effect of tool availability.
\item I translate these ideas into a concrete evaluation protocol, including conditions, measures, item design, timing, and analysis, and use a brief empirical illustration to show why those choices matter.
\end{enumerate}

This paper develops a conceptual framework and proposed protocol. A short empirical illustration in Section~\ref{sec:illustration} draws on a collaborative verification study to clarify the protocol's identification limits; it does not validate the diagnostic categories.

\section{Background and Related Work}

\subsection{What Current Evaluations Tell Us}

Most work on AI-assisted verification answers one of three questions.

First, some studies evaluate the system. Automated fact-checking is often assessed through benchmarks for claim detection, evidence retrieval, and verdict prediction \citep{guo2022survey,thorne2018fever,schlichtkrull2023averitec}. These studies tell us whether the model performs well in isolation. They do not tell us whether the output improves human judgment, or whether it helps users learn anything.

Second, some studies evaluate the treated claim. Correction and debunking studies ask whether exposure to a correction improves belief accuracy for the claim being corrected \citep{walter2020fact}. Studies of labels, provenance signals, and warning cues do something similar: they measure judgment quality while the cue is present. This is useful, but it still does not tell us how people handle new claims later on.

Third, some studies evaluate the human--AI team. This work looks at team accuracy, trust, reliance, and complementary performance while assistance is available \citep{lee2004trust,bansal2021does,vaccaro2024combinations}. Again, this is an important part of the picture, but it mainly tells us what the tool contributes at the point of use.

So the gap is simple. Current evaluations usually tell us whether the system works, whether a cue works, or whether the assisted decision improves. Delayed, novel-claim evaluation addresses an additional endpoint that these immediate measures do not establish.

\subsection{Affordances, Agency, and Enacted Verification}

AI verification interfaces do more than expose information. They also provide cues about machine authority and action possibilities for the user. In HAII--TIME, visible attributes of an AI system can shape judgments through a cue route, whereas actually using an affordance can shape experience through an action route \citep{sundar2020machine}. This distinction matters because a verification button, source panel, or uncertainty display may be available without being used; conversely, an interface can require an action without necessarily increasing a user's felt ownership of the judgment.

The distinction also cautions against a simple equation of ``more user action'' with better learning. User actions impose time and effort costs, and AI systems are valuable partly because they can remove unnecessary work. What matters for retained capability is therefore not action in the abstract, but whether the interaction causes users to perform, rehearse, or newly acquire operations that are useful later. Recent evidence from Liao and Sundar's preregistered experiment illustrates why the layers should be separated: verification cues, forced verification actions, and voluntary verification actions were manipulated as distinct interface conditions, and verification-action conditions reduced trust and perceived credibility in a misleading generative-AI assistant \citep{liao2026chat}. That study concerns immediate trust and credibility rather than delayed transfer. It did not detect a difference in perceived agency, illustrating why required action and agency should be measured separately rather than equated.

The epistemic-transfer framework therefore treats interface architecture, enacted behavior, and subjective experience as analytically distinct. Randomizing an answer-first versus evidence-first interface can identify the effect of that interface package on ETE or TRC. It does not, by itself, establish whether any effect was caused by perceived agency, extra effort, reduced anchoring, source inspection, or another process. Those mechanisms require additional manipulation or appropriately cautious process analysis.

\subsection{Why Retained Capability Matters}

People do not approach information in the same way when tools are available. Cognitive offloading is often useful: external resources can save time and free up mental effort \citep{risko2016cognitive}. At the same time, access to external support can change how much people search, remember, or reason for themselves \citep{fisher2015searching,sparrow2011google}. That is not necessarily bad. But it means we should not assume that strong assisted performance automatically implies stronger independent performance later.

Automation research makes a similar point. Automation can improve safety and efficiency, while also reducing opportunities to maintain human skill \citep{bainbridge1983ironies}. Studies in aviation, for example, show that infrequently practiced flying skills---especially cognitive ones such as navigation and procedure recall---degrade in the automated cockpit \citep{casner2014retention}. Meta-analytic work further shows that the effects of automation depend on how much of the task is automated and at what stage \citep{onnasch2014human}. For AI verification tools, the analogous question is not whether the tool helps now, but what kind of user it helps create over time.

\subsection{Learning, Retention, and Transfer}

Learning research also helps clarify the issue. Doing well during practice is not the same as learning something that lasts \citep{soderstrom2015learning}. Easier practice can improve short-term performance without producing the strongest long-term retention. By contrast, retrieval practice and other desirable difficulties may feel harder in the moment while improving later performance \citep{roediger2006test,bjork2011making}.

Transfer adds another requirement. It is not enough that users remember the exact claim they saw before. What matters is whether they can apply a useful strategy to new claims. Transfer can be near or far depending on how much the later task resembles the earlier one \citep{barnett2002when,salomon1989rocky}. In verification settings, this may mean using lateral reading, checking multiple sources, or paying attention to source expertise \citep{wineburg2019lateral,kozyreva2020citizens}. Notably, some misinformation interventions are explicitly designed with transfer in mind: inoculation-style ``prebunking'' games aim to build generalizable resistance to manipulation techniques rather than to correct individual claims \citep{roozenbeek2019fake}. Transfer is therefore an additional design objective for AI verification tools, rather than an automatic consequence of providing corrections. Whether they nevertheless support transferable strategies---or bypass them---is an empirical question.

\subsection{Emerging Evidence from Generative and Clinical AI}

Recent experiments motivate evaluation after assistance while also showing why task and comparison matter. In high-school mathematics, unrestricted generative AI support improved practice performance but reduced subsequent unassisted performance relative to a no-AI control; a scaffolded tutor largely mitigated that harm \citep{bastani2025generative}. Randomized experiments in mathematics and reading also report lower persistence and poorer immediate unassisted performance after brief AI assistance \citep{liu2026ai}. In programming, randomized AI access reduced assessed mastery of an unfamiliar library, while observed interaction patterns suggest that how developers use assistance may matter for learning \citep{shen2026how}. These results do not by themselves establish durable skill loss or identify engagement as a causal mediator.

Verification studies add a distinction between treated beliefs and novel-claim performance. Imperfect LLM fact-checking guidance can reduce headline discernment while assistance is present \citep{deverna2024fact}. In a four-week dialogue study, Rani et al. found immediate assisted gains but no statistically detectable improvement in pre-assistance accuracy across sessions; accuracy on new items immediately after AI use declined across weeks. The absence of a longitudinal no-AI comparison limits causal interpretation of that decline \citep{rani2026dialogues}. Immediate correction, transfer to new claims, and change over time therefore require separate comparisons.

Evidence from longer-term use is suggestive but less causally direct. A quasi-experimental working paper using 30 months of panel data from 26{,}811 Chinese secondary students estimates higher homework scores but lower closed-book exam scores after AI adoption, with longer-run penalties on entrance examinations \citep{stromberg2026penalty}. Its estimates depend on the assumptions of the staggered-adoption design. In an observational multicentre colonoscopy study, unassisted adenoma detection was lower after AI introduction than before it (22.4\% versus 28.4\%). This before--after association raises a deskilling concern but does not isolate AI exposure from other changes over time \citep{budzyn2025endoscopist}. Self-reported reductions in critical-thinking effort likewise describe experiences rather than demonstrate loss of skill \citep{lee2025impact}.

Other findings show why harm should not be assumed. In a dynamic decision-making task, Karny et al. observed immediate assistance benefits without detecting a subsequent skill decrement or learning carryover \citep{karny2024learning}. In an exercise-recommendation task, contrastive explanations improved immediate independent performance relative to unilateral explanations \citep{bucinca2025contrastive}. Work on incidental learning also shows that the form of assistance matters \citep{gajos2022do}. These studies differ in task, comparator, and timing; none substitutes for a delayed verification test. Together, they motivate an evaluation that can identify benefits, practical equivalence, penalties, and unresolved effects of prior assistance.

\section{Epistemic Transfer}

\subsection{Definition}

I define \emph{epistemic transfer} as the effect of prior interaction with an AI verification system on later performance when evaluating novel claims without that system or equivalent AI assistance, relative to a specified comparison activity.

Three features matter.

First, the later outcome must be measured without the target system or functionally equivalent AI assistance. The access regime must also specify whether participants retain ordinary web search, evidence cards, or other non-AI resources. A source-supported task and a closed-resource test measure different capabilities; both can be unassisted by the target AI. In remote studies, record checks on prohibited outside assistance and the limits of those checks.

Second, the later test must use novel claims. Otherwise we may just be measuring memory for a claim or correction seen earlier.

Third, the later test should happen after a retention interval. Otherwise we risk confusing lasting change with short-lived activation.

Epistemic transfer is also comparison-based. A system can help more than no additional practice, while still helping less than active verification practice. Both comparisons matter. The first estimates the effect of additional AI practice relative to the specified control activity. The second estimates its opportunity cost or benefit relative to active verification practice. Neither comparison alone establishes within-person skill loss from baseline.

Epistemic transfer is an outcome construct, not a mechanism claim. The same ETE could arise through different combinations of enacted verification, perceived agency, effort, feedback, anchoring, or trust. Conversely, an interface can increase perceived agency or interaction without producing measurable transfer. ETE therefore answers what prior use changed in later independent performance; explaining why requires separate process measures or mechanism-identifying manipulations.

This outcome is different from several nearby ideas. A correction effect concerns whether belief accuracy improves for a treated claim \citep{walter2020fact}. Trust and reliance concern how people respond to the system while it is present \citep{lee2004trust}. Human--AI team performance concerns the quality of the joint decision. None of these tells us, by itself, how well the user performs later on a new claim without the tool.

\begin{table}[t]
\centering
\caption{Core conditions in the proposed evaluation protocol.}
\label{tab:conditions}
\small
\begin{tabularx}{\textwidth}{@{}p{3.0cm}YY@{}}
\toprule
\textbf{Condition} & \textbf{What the user experiences} & \textbf{Why it is included} \\
\midrule
Answer-first AI & The system gives a verdict and explanation before the user records a judgment. & Captures an architecture in which the system conclusion can act as a cue or anchor before verification action. \\
\addlinespace
Evidence-first AI & The system presents evidence and requires an initial judgment before revealing the same verdict and explanation. & Tests an architecture that provides and can require verification action before the verdict; it does not by itself isolate agency, effort, or engagement as the mechanism. \\
\addlinespace
Active practice & Participants verify comparable claims without AI, using the same non-AI resources and a specified reference-feedback policy. & Shows what users retain from the verification practice that AI may replace. \\
\addlinespace
No additional practice & Participants complete an unrelated matched-duration activity. Common assessments are still verification exposure. & Estimates the effect of additional practice relative to the specified control activity and assessment regime. \\
\bottomrule
\end{tabularx}
\end{table}

\subsection{Research Questions}

The framework is organized around four practical questions:

\begin{description}
\item[RQ1:] How does prior AI-assisted verification affect delayed unassisted accuracy, calibration, and verification behavior on new claims, compared with active practice and with no additional practice?
\item[RQ2:] How much does immediate performance change with randomized tool availability after different practice interfaces?
\item[RQ3:] Which interface architectures---including answer timing, verification affordances, and required versus voluntary action---increase or reduce epistemic transfer, and which process indicators accompany those effects?
\item[RQ4:] How do these effects vary by transfer distance, baseline skill, domain knowledge, and language or media context?
\end{description}

A single study need not answer all four questions. It should state which contrasts it identifies: omitting a comparator narrows the conclusions, and subgroup or transfer-distance claims require enough participants and distinct items to support them.

\section{Two Complementary Estimands}
\label{sec:estimands}

Both quantities are ordinary causal contrasts organized around different moments of evaluation, using potential outcomes to state their targets \citep{hernan2020causal}. Their value lies in specifying what is compared, under which exposure and access regime, and then reporting them together. ETE changes the earlier practice assignment while holding the later test regime fixed; TRC changes current availability while holding prior history fixed. For clarity, the definitions below use binary accuracy, where larger values mean better performance; other outcomes need their own scale and direction.

\subsection{Epistemic Transfer Effect}

Let $P$ be the target participant population and $Q_d$ the distribution of novel claims at transfer distance $d$. Let $Y^{\mathrm{delay}}_{ij}(c,q;\tau,b)$ denote participant $i$'s potential accuracy on claim $j$ after assignment to practice condition $c$, under assessment policy $q$, at delay $\tau$, with access regime $b$. Policy $q$ specifies intervening assessments, feedback, and any probe-availability randomization, including how these depend on practice condition. Define
\begin{equation}
\mathrm{ETE}(c,k;\tau,d,b,q)
=\mathbb{E}_{i\sim P,j\sim Q_d}\!\left[
Y^{\mathrm{delay}}_{ij}(c,q;\tau,b)-Y^{\mathrm{delay}}_{ij}(k,q;\tau,b)
\right].
\label{eq:ete}
\end{equation}
The expectation also averages over any randomization in $q$. ETE asks how prior assignment changes later performance relative to comparator $k$, for the same target population, items, delay, and access regime. Baseline covariates can improve estimation; they do not replace specification of the population over which the contrast is averaged.

Random assignment identifies an intention-to-treat contrast for the enrolled population under consistency, no interference between participants, and adequate outcome observation or defensible missing-data assumptions. Generalization to another population or claim universe requires additional sampling or transport assumptions. If $q$ includes condition-specific probe assistance, ETE is the effect of the complete sequence. ETE under $q=\text{no probe}$ is a different estimand.

Report ETE separately against active practice and no additional practice. Equivalence to active practice is compatible with both groups learning. A negative ETE against active practice indicates lower delayed performance than that alternative, not necessarily deterioration from baseline. A negative effect against no additional practice strengthens a relative-harm interpretation, but absolute change still requires comparable measurement across time.

\subsection{Tool-Removal Cost}

Let $A$ indicate current target-tool availability, and let $H$ summarize exposure before a probe trial after practice condition $c$. For a specified probe policy $\pi$, define
\begin{equation}
\mathrm{TRC}(c;\pi,b)
=\mathbb{E}_{P,Q_{\mathrm{probe}},H\sim\pi_c}\!\left[
Y^{\mathrm{probe}}_{ij}(c,H,1;b)-Y^{\mathrm{probe}}_{ij}(c,H,0;b)
\right].
\label{eq:trc}
\end{equation}
Here $b$ specifies the non-AI resources held constant; $A$ toggles the target tool. The contrast averages over the same joint participant, eligible-item, and prior-history distribution induced by the probe policy in its two terms. Identification requires randomized current availability with positive assignment probabilities in the histories to which inference applies, plus consistent exposure and outcome measurement. A first-trial probe avoids within-probe history; a repeated probe targets the histories generated by its specified policy. Designs without the needed support should define and estimate a sequence-level contrast instead.

TRC measures a current availability effect. Positive values indicate an advantage from the tool; negative values indicate worse current performance with it. The label ``cost'' is retained for continuity, but an availability contrast is not necessarily a chronological drop after abrupt withdrawal, a measure of dependence, or evidence that prior skill has been lost. Different-item trials within a person improve efficiency; they do not observe both potential outcomes for that person on the same claim. The primary estimand is an average, not an identified individual causal effect.

TRC is related to human--AI team performance and complementarity \citep{bansal2021does,vaccaro2024combinations}. It is a post-practice human-with-tool versus human-without-tool contrast. Establishing that a team also exceeds the AI alone requires an AI-only benchmark, which TRC does not supply. A large TRC can accompany useful retained capability, no relative transfer advantage, or a transfer penalty. The joint reading below keeps these possibilities separate.

\section{Proposed Evaluation Methodology}

\subsection{Overview}

\begin{figure}[!htbp]
\centering
\includegraphics[width=\textwidth]{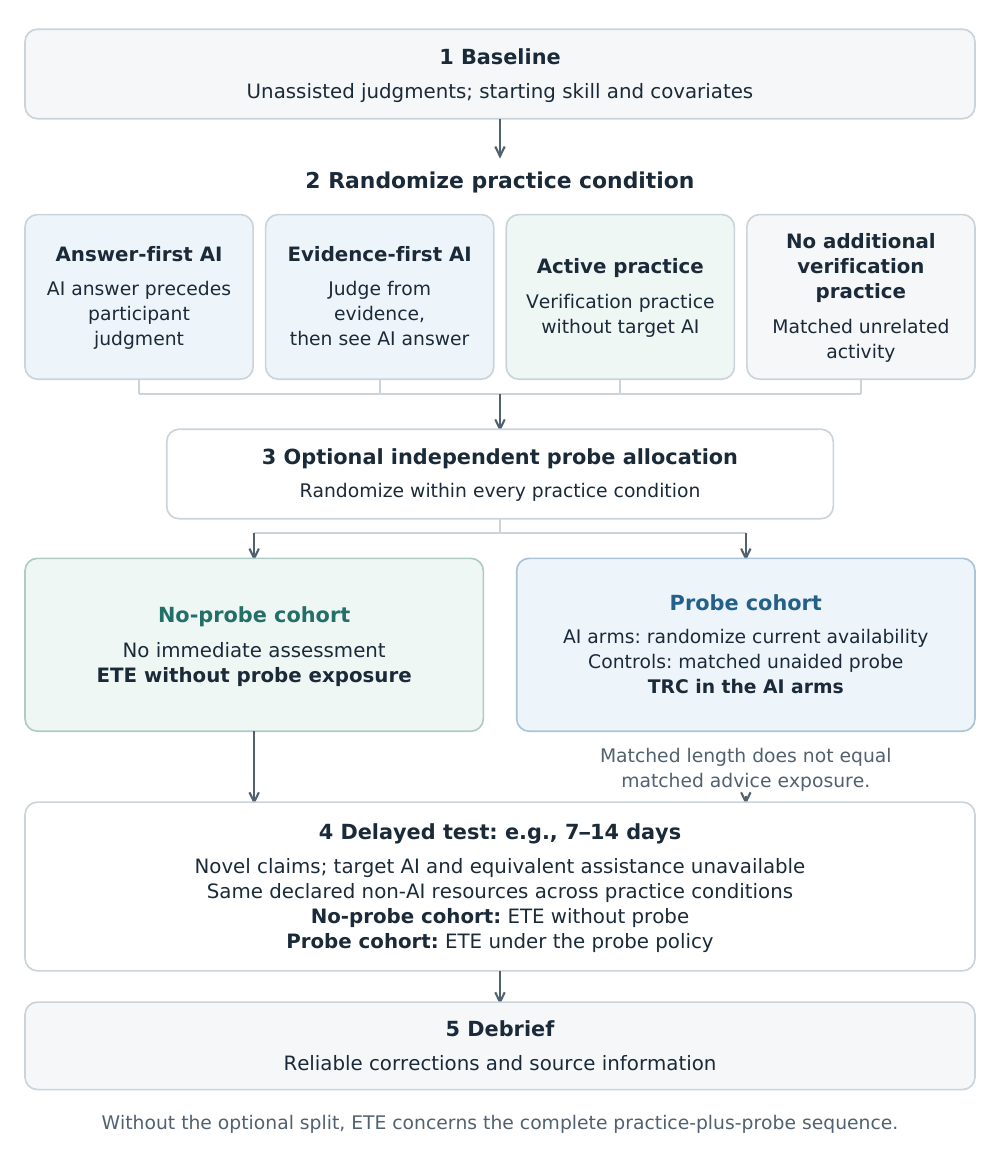}
\caption{Proposed protocol. Practice condition is randomized after baseline. A second, independent allocation to probe or no-probe assessment is optional: it allows delayed ETE to be estimated without additional probe exposure, while TRC is estimated in the probe cohort. Without this split, ETE refers to the complete practice-plus-probe sequence. All delayed tests use novel claims and the same declared non-AI access regime.}
\label{fig:flow}
\end{figure}

The core design randomizes participants to four practice conditions (Table~\ref{tab:conditions}) and later assesses all groups without the target AI. An immediate availability probe adds a second randomization within the AI conditions. Figure~\ref{fig:flow} shows the protocol and an optional split that separates the probe from the main delayed assessment.

Randomizing answer-first versus evidence-first identifies the effect of an interface package. To isolate answer timing, hold verdicts, rationales, source access, feedback repetition, judgment opportunities, and revision opportunities constant; otherwise report these differences as part of the treatment. A required source choice or initial judgment does not prove that the participant read the evidence first. Mechanism studies can independently manipulate verification cues, required or voluntary actions, or explanation types \citep{sundar2020machine,liao2026chat,bucinca2021trust}, while distinguishing assignment from observed use.

Specify whether assistance comes from a live model, frozen model outputs, or a scripted reference assistant. Archive model versions, prompts, outputs, retrieval snapshots, and error rates where possible. A correct-only assistant tests interaction with correct advice; it cannot establish appropriate reliance on fallible systems. If error detection matters, vary advice correctness independently of interface and item difficulty in safe tasks and report responses to correct and incorrect advice separately.

\subsection{Participants and Sampling}

Participants should match the intended users of the system. A public-facing tool should be tested with a broad adult sample, not only with students or technically trained users. A professional tool should be tested with the relevant professionals. Multi-country or multilingual studies are especially useful when model quality, source availability, or information environments differ across contexts.

Sample size should be based on the smallest effect that matters in the delayed test. Because repeated responses are crossed by participants and claims, simulation-based power analysis is preferable to a simple independent-samples calculation. Eligibility rules, exclusions, attrition handling, and attention checks should be preregistered.

\subsection{Claims, Evidence, and Transfer Distance}

The claim pool should include true and false claims and, when relevant, mixed or uncertain ones. Claims should come from domains that users plausibly encounter, such as health, politics, climate, or provenance. Each claim should have a documented reference judgment grounded in high-quality evidence.

Create disjoint item sets within each participant for baseline, practice, probe, and delayed assessment. Where feasible, rotate matched item lists across phases between participants, so phase is not confounded with a single fixed claim set. Randomize probe-item assignment to availability as well as order. Document reference sources, adjudication disagreements, truth-label counts, and the weighting of those labels in the target outcome. Pilot for ceiling and floor effects and for adequate evidence. A pre/post difference on distinct, uncalibrated item sets is not a measure of absolute learning or skill loss.

Novelty should be treated carefully. Following transfer research \citep{barnett2002when}, delayed-test claims can be grouped into near, intermediate, and far transfer. Near-transfer claims resemble the practice claims in domain and evidence structure. Intermediate-transfer claims stay in the same domain but require a different kind of evidence. Far-transfer claims require the same verification strategy in a different domain. Textual and semantic overlap checks, together with human review, can help prevent paraphrases of practiced claims from slipping into the delayed test. Truth status and transfer distance are separate dimensions: a misleading claim is not necessarily far transfer. Include multiple distinct claims in each distance-by-domain category; repeated judgments of one claim do not create additional item replication.

\subsection{Procedure}

\textbf{Phase 1: Baseline.} Collect judgments on novel-to-participant claims without AI under the declared resource regime. Elicit confidence with explicit wording, and measure baseline verification behavior if later behavior--outcome associations are planned. Keep baseline procedures common across conditions; baseline itself is verification exposure.

\textbf{Phase 2: Randomized practice.} Assign participants to one of the four conditions. Choose the amount and spacing of practice to match the intended use and learning opportunity. Match or explicitly vary task duration, evidence, reference feedback, and opportunities for judgment and revision. The no-additional-practice condition completes a matched-duration unrelated activity.

\textbf{Phase 3: Immediate availability probe.} In AI conditions, independently randomize current assistance on novel items under a prespecified assignment policy. Balance item difficulty across availability conditions through assignment, not merely by matching average scores after collection. Record trial order, prior availability, and feedback. Fresh claims prevent simple repetition but do not remove strategy learning or carryover from earlier assisted trials. A first-probe-trial contrast estimates availability before within-probe carryover; a mixed-trial contrast estimates an average under its exposure history. They answer different questions.

If controls also complete a matched-length unassisted probe, exposure count is comparable but learning opportunities remain different: AI arms see additional advice and controls complete more unassisted judgments. Consequently, delayed ETE estimates the effect of the complete sequence under this assessment policy. To target practice effects without a probe, independently randomize probe participation within every practice condition and reserve the no-probe cohort for the corresponding delayed ETE. The probe cohort can still complete follow-up to estimate how probing changes the practice contrast. A smaller study may omit the probe and estimate ETE alone.

\textbf{Phase 4: Delayed test.} After a prespecified interval, for example 7--14 days, administer novel claims with the target AI and equivalent assistance unavailable. Specify the allowed non-AI resources, follow-up window, and handling of late responses. Add longer follow-ups when repeated professional or educational use is the target; a one-week test does not establish durable effects over months. Use distinct or counterbalanced item lists across follow-ups. A later test after an earlier follow-up targets a repeated-assessment policy; a test at the same delay without that earlier follow-up is a different regime.

\textbf{Phase 5: Debriefing.} Provide reliable corrections and source information. If prompt correction is needed before the last follow-up, treat it as another exposure, standardize it where feasible, and include its timing in the assessment policy. Do not postpone necessary correction merely to preserve an uncontaminated outcome.

\subsection{Outcome Measures}

Choose a primary outcome tied to the intended capability. The following measures answer complementary questions.

\textbf{Accuracy and discernment.} The main outcome can be binary accuracy, an ordinal veracity judgment, or a continuous credibility rating. For balanced true and false claims, discernment can be defined as the difference in mean credibility assigned to true versus false claims.

\textbf{Confidence and calibration.} If probabilistic calibration is an outcome, ask for the probability that the selected answer is correct, give explicit anchors and training, and state the scoring rule. For elicited probabilities $p_{ij}\in[0,1]$, the mean of $(p_{ij}-Y_{ij})^2$ is a Brier score for answer correctness; report accuracy and calibration summaries alongside it because this score also reflects discrimination \citep{brier1950verification}. Alternatively, elicit a probability distribution over the possible veracity labels and use a multiclass score. A generic ``guessing''--``certain'' confidence slider is an ordinal or numerical rating, not automatically a probability scale. Rescaling it to $[0,1]$ does not validate probabilistic calibration or an overconfidence claim.

\textbf{Verification behavior.} Behavioral measures can include search queries, opened sources, source diversity, and logged comparison actions. Distinguish access or dwell time from comprehension, and distinguish selecting a designated source card from open-web verification. Code lateral reading or inspection of conflicting evidence from observable records with an explicit rubric \citep{wineburg2019lateral}.

\textbf{Effort and persistence.} Time on task, number of search actions, abandonment, and willingness to continue after difficulty can describe effort and persistence. Time alone does not establish cognitive engagement; report it alongside accuracy and distinguish time to initial and final judgments \citep{liu2026ai,lee2025impact}.

\textbf{Process and mechanism indicators.} Studies should record whether a verification affordance was available, whether action was required or voluntary, whether the participant actually used it, and what evidence was inspected. Perceived control, ownership of the judgment, mental effort, perceived system authority, and perceived outsourcing can help characterize the interaction. Use validated multi-item measures where available. An item asking whether users formed a judgment before seeing advice records their report of the sequence; it is not by itself a measure of agency or ownership. These distinctions separate an action opportunity from enacted action and from felt agency \citep{sundar2020machine,liao2026chat}. Unless the design directly randomizes the mechanism, however, post-treatment process variables should be treated as descriptive or exploratory rather than as proof of causal mediation.

\textbf{Secondary outcomes.} Trust, usefulness, satisfaction, reliance, and intention to reuse are still worth measuring, but they should not be confused with epistemic transfer.

\subsection{Analysis Plan}

Specify the primary comparator, outcome, delay, and smallest effect of interest before collection. Comparing each of two AI interfaces with each of two comparators yields four delayed contrasts, before adding outcomes or follow-ups. Designate a primary contrast or define the confirmatory family and multiplicity procedure. Keep later sensitivities and mechanism searches explicitly exploratory.

For binary delayed accuracy, one possible model is
\begin{equation}
\operatorname{logit}\Pr(Y_{ij}=1\mid u_i,v_j)
=\beta_0+\boldsymbol\beta_C^{\top}C_i+\boldsymbol\beta_D^{\top}D_j
+C_i^{\top}B D_j+\boldsymbol\gamma^{\top}X_i+u_i+v_j,
\label{eq:model}
\end{equation}
where $C_i$ and $D_j$ are vectors coding condition and transfer distance, $X_i$ contains baseline covariates, and participant and claim effects are crossed. For sufficiently replicated designs, consider claim-specific condition slopes and, in the availability model, participant-specific availability slopes. Preregister a convergence and singularity fallback; random intercepts alone do not represent all treatment-effect variation. If inference is limited to a fixed claim bank, state that restriction.

Estimate ETE and TRC on a common response scale, preferably percentage-point accuracy differences with uncertainty intervals. Average response-scale predictions over the same prespecified participant and item distribution for each condition. Report how random effects are handled: integrating over their distribution targets a population-average prediction, whereas setting them to zero gives a conditional prediction. Averaging logits and then applying the inverse logit is also different from averaging probabilities. Neither a default reference grid nor a probability at mean covariates should be silently described as population standardization. Raw group rates remain useful descriptive checks. Report assisted and unassisted probe means alongside TRC: a larger difference can reflect a higher assisted mean, a lower unassisted mean, or both. Uncertainty should represent both participants and claims when generalizing beyond the sampled item bank, and retain dependence between contrasts from the same participants.

For TRC, include availability, interface, their interaction, item and order structure, and any history terms justified by the randomization policy. Use that policy in estimation, including known assignment probabilities if these vary. Probe-history adjustment is not a substitute for randomization or adequate support. Do not fit person-specific ``transfer effects'' by treating individual observed scores as counterfactual contrasts; TRC--ETE associations across a small number of interfaces are also insufficient to establish a general law.

A nonsignificant test is inconclusive about practical equivalence. Prespecify bounds on the reported response scale, then use equivalence tests or interval decisions matched to the intended claim \citep{lakens2017equivalence}. At a two-one-sided-tests level of .05, an unadjusted 90\% confidence interval must lie wholly within the equivalence bounds. Report 95\% intervals for effect estimation, and account for any planned family of decisions rather than selecting the most favorable interval after analysis.

\textbf{Attrition and missingness.} Define the primary target as the mean effect of assignment in the randomized cohort, including participants with unobserved delayed outcomes. Report randomized and observed counts by condition, follow-up timing, missing items, and baseline predictors of return. Complete-case contrasts do not retain randomization automatically. Excluding post-treatment failures of speed, attention, or assistance checks can also change the target; distinguish assignment analyses from prespecified compliance sensitivities. Prespecify imputation or weighting under explicit missing-at-random assumptions, and sensitivity analyses for plausible outcome-dependent dropout. Report how conclusions change; covariate balance or similar retention rates alone cannot verify those assumptions.

\textbf{Process analyses.} Treat source inspection, confidence, effort, agency, and trust as distinct outcomes or candidate mediators. Include relevant baseline behavior when examining incremental associations, while acknowledging measurement error and residual confounding. Conditioning on post-treatment variables can block treatment pathways or introduce bias. Even a persistent adjusted association neither identifies causal mediation nor establishes predictive validity on new participants or claims; mechanism claims need additional design and assumptions.

\subsection{Feasibility and Precision Planning}

There is no defensible default sample size for this protocol without a specified effect, item distribution, variance structure, and decision rule. Simulate the planned randomization, participant and item heterogeneity, expected accuracy, label mix, dropout, and analysis. Vary both participants and distinct claims: adding ratings cannot compensate fully for too few items when generalization across claims matters \citep{westfall2014power}. Include uncertainty in the pilot variance estimates and document Monte Carlo error and convergence failures.

Power a superiority claim, an equivalence claim, and an availability-by-interface interaction for their own targets; one calculation does not cover all three. Report the assumed effects and margins, recruitment and retained sample sizes per arm, analysis code, and the range of precision across plausible scenarios. A split probe design also needs enough participants in each assessment cohort. If resources are limited, reduce the number of confirmatory contrasts or treat the study as estimation with acknowledged uncertainty.

Plan recruitment using an explicit retention scenario: if $n$ delayed observations per arm are required and expected retention is $r$, $\lceil n/r\rceil$ recruits per arm is an expectation-based starting point before other exclusions. It does not guarantee $n$ observed outcomes; buffers should reflect uncertainty when a minimum completer count is required. For example, 80\% retention implies recruiting 25\% above the target; this is arithmetic, not evidence that 80\% is the correct assumption for a particular panel. Retention estimates should come from a relevant pilot or closely matched design. No power simulation or universal sample-size recommendation is claimed here.

\section{Diagnostic Space}

\begin{figure}[!htbp]
\centering
\includegraphics[width=\textwidth]{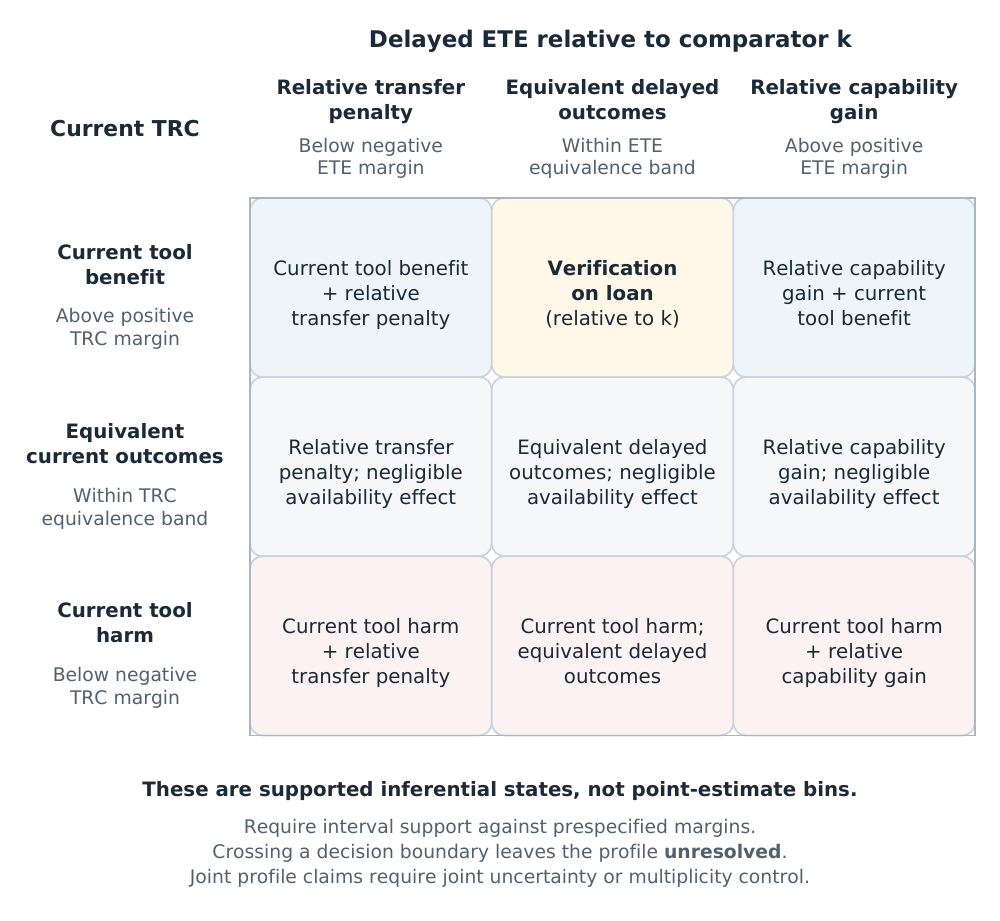}
\caption{Interpretation matrix for ETE and TRC, conditional on a named comparator and assessment regime. Columns and rows are supported inferential states, not bins for point estimates. Margins are prespecified; intervals that do not support one of these states remain unresolved. ``Verification on loan'' is comparator-relative. The matrix includes negative availability effects and separates equivalence from a relative transfer penalty.}
\label{fig:space}
\end{figure}

Read ETE and TRC jointly for a named comparator, delay, access regime, and item distribution (Figure~\ref{fig:space}). Specify practical margins $\delta_E$ and $\delta_A$ on their response scales before observing results. Report the estimates and uncertainty region first. Meaningful superiority, meaningful inferiority, and equivalence require their own interval decisions; an interval crossing a decision boundary leaves the corresponding profile unresolved.

\textbf{Relative capability gain.} ETE is meaningfully positive against the named comparator. If TRC is also meaningfully positive, the profile is \emph{relative capability gain plus current tool advantage}; if TRC is practically equivalent to zero, the delayed advantage occurs without a material current availability effect. These labels concern measured performance and do not identify a learning mechanism. The label ``gain'' is relative to the comparator, not a within-person change score.

\textbf{Verification on loan, relative to a comparator.} TRC is meaningfully positive while delayed ETE is practically equivalent to zero against that comparator. The extra current assistance benefit coexists with equivalent delayed performance. Against active practice, this does not mean that nothing was learned: both conditions may outperform no additional practice. If delayed ETE is meaningfully negative, describe the stronger pattern as \emph{current tool advantage with a relative transfer penalty}. Do not pool equivalence, inferiority, and imprecision into a single ``no transfer'' category.

\textbf{Equivalent measured outcomes or relative penalty.} When both ETE and TRC are practically equivalent to zero, delayed performance is equivalent to the named comparator and current tool availability has no material effect within the AI condition. A meaningfully negative ETE identifies a relative delayed penalty regardless of TRC. Neither warrants the global label ``epistemically inert'' or ``de-skilled'' without further evidence.

\textbf{Current tool harm.} A meaningfully negative TRC means the available tool worsens immediate performance in the tested regime. This can coexist with any delayed ETE and must remain visible in the framework, especially with fallible advice.

A hypothetical example illustrates why both comparators matter. Suppose delayed accuracy, averaged over the same target population and claim distribution, is 70\% after AI practice, 70\% after active practice, and 60\% after no additional practice, with TRC of 20 percentage points. ETE is then 0 points against active practice and $+10$ against no additional practice. If uncertainty supports the corresponding equivalence and superiority decisions, the same interface has a verification-on-loan profile relative to active practice and a capability-gain profile relative to no additional practice. It has not ``left nothing behind.'' These numbers are illustrative, not study results.

If profiles depend on decisions for both estimands, report joint uncertainty using simultaneous intervals or a joint region, or control the planned decision family. Separate unadjusted significance tests should not be presented as a calibrated joint classification. The matrix is an interpretive aid rather than a universal certification score.

\clearpage
\section{An Initial Empirical Illustration}
\label{sec:illustration}

This section summarizes selected results from the same collaborative two-wave health-claim verification study analyzed in a separate empirical manuscript; it is not an independent replication. The study randomized 552 U.S. adults to answer-first AI, evidence-first AI, or source-practice, with ten practice claims. A total of 474 participants returned after a median of eight days to judge sixteen novel claims with evidence cards but no AI answer. The assistant displayed fixed, correct reference judgments. The conditions differed in sequence, initial commitment, revision, and feedback exposure, rather than answer timing alone.

The immediate probe illustrates a design limitation. In both AI conditions, the first four probe trials were assisted and the last four were unassisted; claim order was shuffled. Observed accuracy was 88.6\% versus 69.2\% in answer-first and 92.1\% versus 64.3\% in evidence-first, giving early--late gaps of 19.4 and 27.7 percentage points. Availability was confounded with period and potentially with practice, fatigue, or carryover. These descriptive gaps therefore do not identify the randomized TRC in Equation~\ref{eq:trc}; shuffled claim order does not resolve that confounding.

The baseline-adjusted delayed comparison did not detect an evidence-first advantage over answer-first (logistic mixed-model odds ratio $0.791$, 95\% CI $[0.624,1.002]$, $p=.052$). The response-scale equivalence analysis with five-percentage-point bounds was also inconclusive. This comparison concerns assignment to the full interaction sequence among observed returners. Without a no-additional-practice control, and with claim sets fixed by phase, it does not establish absolute learning or skill loss.

The team records a prespecified protocol, but the study was not externally registered. This illustration motivates independent availability randomization and explicit comparators; it does not establish a ``verification-on-loan'' classification or a psychological mechanism. The broader empirical analyses are outside this short illustration.

\section{Implications for Research, Design, and Policy}

\subsection{Research}

The framework gives researchers a common language for comparing studies that currently use different concepts and endpoints. At a minimum, studies should report the comparator, delay, access regime, item novelty, transfer distance, and outcome family alongside any transfer estimate. Without these details, two studies may appear to disagree while in fact estimating different things.

The framework also makes heterogeneity an explicit research question. Effects may differ by baseline skill, age, domain expertise, language, or media environment. Some users may benefit most from immediate answers, while also losing the most opportunity to practice verification.

\subsection{System Design}

The design question is not just whether to add explanations. It is what cues and action possibilities the interface creates, which of those actions users actually enact, and what downstream consequences follow. Answer-first interfaces, evidence-first interfaces, verification cues, forced or voluntary checking, cognitive forcing, and uncertainty displays can all be compared using the same ETE and TRC outcomes \citep{sundar2020machine,liao2026chat}. Existing work suggests that engagement and workflow cost shape overreliance and incidental learning \citep{bucinca2021trust,gajos2022do,vasconcelos2023explanations}.

This also makes trade-offs visible. More user action is not intrinsically better: verification actions consume time and effort, and a system may rationally automate operations that do not need to be retained. A more demanding interface may reduce speed or satisfaction while improving retained capability, while a very fluent interface may maximize short-term performance with little transfer. ETE and TRC make these trade-offs measurable rather than presuming that either convenience or activity is the correct design objective.

\subsection{Policy and Procurement}

The EU AI Act addresses AI literacy, transparency, and, for high-risk systems, human oversight \citep{euaiact2024}. Retained independent judgment can inform research on effective oversight, but the proposed transfer measures are an evaluation recommendation, not a claimed legal requirement. They are particularly relevant when organizations claim that an AI system supports human competence or informed judgment.

A simple procurement checklist follows from the framework. Decision makers can ask: Does the system improve immediate decisions? How does performance change when the system is unavailable? Does repeated use improve or weaken later independent performance? The observational colonoscopy study illustrates the need to monitor unassisted performance as well as point-of-use outcomes, while also testing alternative explanations for changes over time \citep{budzyn2025endoscopist}. Different answers may call for different responses, including interface redesign, more realistic claims about the tool, fallback procedures, or periods of unassisted practice.

\section{Boundary Conditions and Limitations}

Epistemic transfer is not equally relevant for every tool. It matters most when use is repeated, the task contains learnable strategies, users are likely to face similar situations without equivalent assistance, and errors have meaningful consequences. A calculator can be valuable without teaching arithmetic. Likewise, a specialized professional tool may be useful even if users cannot match its full performance independently.

The proposed measures also have limits. Novel claims may still resemble practiced claims in ways that are hard to detect. Transfer distance is domain-specific. Process measures can change when the study constrains normal search behavior. The removal probe can itself teach, expose participants to further advice, or change their expectations. Its effect on delayed contrasts may attenuate, amplify, or reverse them; the direction is not known from matched probe length. Independently randomized probe participation can separate these regimes, at a cost in sample size. The diagnostic space should therefore not be reduced to a single universal certification score.

A further limitation concerns mechanism identification. An answer-first versus evidence-first contrast typically changes several things at once: when an authoritative machine conclusion appears, whether verification action occurs before that conclusion, how much effort is required, and potentially the user's perceived agency or ownership. A treatment effect on ETE identifies a contrast between interface packages, not a pure claim that ``agency'' or ``epistemic work'' alone caused it. Researchers who need that stronger inference should factorially manipulate the relevant affordance and its enactment---for example, cue versus forced versus voluntary verification action---or use other controls that match initiation and effort while reallocating the target operation \citep{sundar2020machine,liao2026chat}. Such manipulations can discriminate candidate accounts, but claims about psychological mediators still require assumptions about what else the manipulation changes.

Tool-removal probes also require judgment. Assistance should not be withheld during real, high-stakes decisions if participants would ordinarily be entitled to use it. In such cases, removal is better studied through simulations, retrospective tasks, or other low-risk exercises. Finally, this framework does not validate the proposed profile labels or thresholds. Predictive value, measurement reliability, and usefulness to design decisions require prospective evaluation across domains. Its terminology should complement established learning and transfer research rather than replace it.

\section{Conclusion}

AI-assisted verification should be evaluated at two distinct moments when later independent judgment matters. ETE estimates what prior assignment changes in delayed performance relative to a named alternative. TRC estimates what current tool availability contributes after practice. Neither alone establishes learning, dependence, or de-skilling.

The proposed protocol makes those contrasts explicit through active and no-additional-practice comparators, novel claims, declared access and assessment regimes, randomized availability, and uncertainty-aware analysis. The empirical illustration shows that a promising study structure does not guarantee identification: an ordered assistance schedule and an inconclusive delayed comparison must remain qualified in the conclusions. It distinguishes the interface that is assigned from actions that are observed and mechanisms that remain to be tested. A credible evaluation should be able to report a benefit, a penalty, practical equivalence, or an unresolved result without changing its standards after seeing the data. The aim is to make claims about what assistance leaves behind as testable as claims about what it helps users do now.

\section*{Declarations}

\textbf{Funding.} This work was supported in part by the Research Council of Norway through SFI MediaFutures, Research Centre for Responsible Media Technology and Innovation (grant no.\ 309339).

\textbf{Competing interests.} The author declares no competing interests.

\textbf{Empirical scope.} Section~\ref{sec:illustration} summarizes selected outcomes from an existing collaborative study. Participants provided informed consent, and the study was documented through the institutional research and data-protection process. This documentation is not presented as independent ethics-board approval. The full empirical account is separate from this framework paper.

\textbf{AI-assisted drafting.} Generative AI tools, including OpenAI Codex during this revision, supported manuscript drafting and editing, literature and methodological checks, and figure-generation code. The author retains responsibility for the arguments, references, and final manuscript. This revision adds a brief summary of an existing collaborative study; it reports no newly collected participant data. Aggregate values used in the illustration accompany the source package.

\bibliographystyle{plainnat}
\bibliography{references}

\end{document}